\documentstyle[12pt]{article}
\catcode`\@=11

\newcommand{\be}{\begin{equation}}
\newcommand{\ee}{\end{equation}}
\newcommand{\bea}{\begin{eqnarray}}
\newcommand{\eea}{\end{eqnarray}}

\def\darr#1{\raise1.5ex\hbox{$\leftrightarrow$}\mkern-16.5mu #1}
\def\part{\partial}

\def\m\mu
\def\n{\nu}

\def\@normalsize{\@setsize\normalsize{15pt}\xiipt\@xiipt
\abovedisplayskip 14pt plus3pt minus3pt%
\belowdisplayskip \abovedisplayskip
\abovedisplayshortskip  \z@ plus3pt%
\belowdisplayshortskip  7pt plus3.5pt minus0pt}
\def\small{\@setsize\small{13.6pt}\xipt\@xipt
\abovedisplayskip 13pt plus3pt minus3pt%
\belowdisplayskip \abovedisplayskip
\abovedisplayshortskip  \z@ plus3pt%
\belowdisplayshortskip  7pt plus3.5pt minus0pt
\def\@listi{\parsep 4.5pt plus 2pt minus 1pt
            \itemsep \parsep
            \topsep 9pt plus 3pt minus 3pt}}

\def\underline#1{\relax\ifmmode\@@underline#1\else
        $\@@underline{\hbox{#1}}$\relax\fi}
\@twosidetrue
\relax

\catcode`@=12

\catcode`\@=11

\def\section{\@startsection{section}{1}{\z@}{3.5ex plus 1ex minus
   .2ex}{2.3ex plus .2ex}{\large\bf}}

\def\ps@headings{\def\@oddfoot{}\def\@evenfoot{}
\def\@oddhead{\hbox{}\hfill
        \makebox[.5\textwidth]{\raggedright\ignorespaces --\thepage{}--
        \hfill }}
\def\@evenhead{\@oddhead}
\def\subsectionmark##1{\markboth{##1}{}} }
\renewcommand{\subsection}[1]{\addtocounter{subsection}{1}
\vspace{2.5mm}\par\noindent {\em \thesubsection . #1}\par
 \vspace{0.5mm} }

\ps@headings

\catcode`\@=12

\relax

\def\figcap{\section*{Figure Captions\markboth
        {FIGURECAPTIONS}{FIGURECAPTIONS}}\list
        {Fig. \arabic{enumi}:\hfill}{\settowidth\labelwidth{Fig. 999:}
        \leftmargin\labelwidth
        \advance\leftmargin\labelsep\usecounter{enumi}}}
 \relax
\def\tablecap{\section*{Table Captions\markboth
        {TABLECAPTIONS}{TABLECAPTIONS}}\list
        {Table \arabic{enumi}:\hfill}{\settowidth\labelwidth{Table
999:}
        \leftmargin\labelwidth
        \advance\leftmargin\labelsep\usecounter{enumi}}}
 \relax
\def\reflist{\section*{References\markboth
        {REFLIST}{REFLIST}}\list
        {[\arabic{enumi}]\hfill}{\settowidth\labelwidth{[999]}
        \leftmargin\labelwidth
        \advance\leftmargin\labelsep\usecounter{enumi}}}
 \relax

\catcode`\@=11

\def\marginnote#1{}
\newcount\hour
\newcount\minute
\newtoks\amorpm
\hour=\time\divide\hour by60
\minute=\time{\multiply\hour by60 \global\advance\minute by-
\hour}
\edef\standardtime{{\ifnum\hour<12 \global\amorpm={am}%
    \else\global\amorpm={pm}\advance\hour by-12 \fi
    \ifnum\hour=0 \hour=12 \fi
    \number\hour:\ifnum\minute<100\fi\number\minute\the\amorpm}}
\edef\militarytime{\number\hour:\ifnum\minute<100\fi\number\minute}
\def\draftlabel#1{{\@bsphack\if@filesw {\let\thepage\relax
  \xdef\@gtempa{\write\@auxout{\string
    \newlabel{#1}{{\@currentlabel}{\thepage}}}}}\@gtempa
    \if@nobreak \ifvmode\nobreak\fi\fi\fi\@esphack}
     \gdef\@eqnlabel{#1}}
\def\@eqnlabel{}
\def\@vacuum{}
\def\draftmarginnote#1{\marginpar{\raggedright\scriptsize\tt#1}}
\def\draft{\oddsidemargin -.5truein
        \def\@oddfoot{\sl preliminary draft \hfil
        \rm\thepage\hfil\sl\today\quad\militarytime}
        \let\@evenfoot\@oddfoot \overfullrule 3pt
        \let\label=\draftlabel
        \let\marginnote=\draftmarginnote

\def\@eqnnum{(\theequation)\rlap{\kern\marginparsep\tt\@eqnlabel}%
\global\let\@eqnlabel\@vacuum}  }
\def\preprint{\twocolumn\sloppy\flushbottom\parindent 1em
        \leftmargini 2em\leftmarginv .5em\leftmarginvi .5em
        \oddsidemargin -.5in    \evensidemargin -.5in
        \columnsep 15mm \footheight 0pt
        \textwidth 250mmin      \topmargin  -.4in
        \headheight 12pt \topskip .4in
        \textheight 175mm
        \footskip 0pt

\def\@oddhead{\thepage\hfil\addtocounter{page}{1}\thepage}
        \let\@evenhead\@oddhead \def\@oddfoot{} \def\@evenfoot{}  }
\def\titlepage{\@restonecolfalse\if@twocolumn\@restonecoltrue\onecolumn
     \else \newpage \fi \thispagestyle{empty}\c@page\z@
        \def\thefootnote{\fnsymbol{footnote}} }
\def\endtitlepage{\if@restonecol\twocolumn \else  \fi
        \def\thefootnote{\arabic{footnote}}
        \setcounter{footnote}{0}}  
\catcode`@=12
\relax

\def\ps@headings{\def\@oddfoot{}\def\@evenfoot{}
\def\@oddhead{\hbox{}\hfill
        \makebox[.5\textwidth]{\raggedright\ignorespaces --\thepage{}--
        \hfill }}
\def\@evenhead{\@oddhead}
\def\subsectionmark##1{\markboth{##1}{}} }

\ps@headings

\relax

\long\def\@caption#1[#2]#3{\par\addcontentsline{\csname
  ext@#1\endcsname}{#1}{\protect\numberline{\csname
  the#1\endcsname}{\ignorespaces #2}}\begingroup
    \small
    \@parboxrestore
    \@makecaption{\csname fnum@#1\endcsname}{\ignorespaces #3}\par
  \endgroup}
\catcode`@=12

\input epsf

\def\firstpage#1#2#3#4#5#6{
\begin{document}


\begin{titlepage}
\nopagebreak
\title{\begin{flushright}
        \vspace*{-1.8in}
        {\normalsize CERN-PH-TH/2005-267 }\\[-10mm]
        {\normalsize CPHT-RR-074.1205}\\[-10mm]
        {\normalsize LPT-ORSAY-05/86}\\[-10mm]
        {\normalsize INR/TH-26-2005}\\[-10mm]
        {\normalsize hep-th/0512276}\\[-4mm]
\end{flushright}
\vfill {#3}}
\author{\large #4 \\[1.0cm] #5}
\maketitle
\vskip -9mm
\nopagebreak
\begin{abstract} {\noindent #6}
\end{abstract}
\vfill
\begin{flushleft}
\rule{16.1cm}{0.2mm}\\[-4mm]
$^{\dagger}${\small   Unit{\'e} mixte de recherche du CNRS et de l'EP,
UMR 7644 . }\\[-4mm]
$^{\ddagger}${\small Unit{\'e} mixte de recherche du CNRS, UMR 8627 .}\\
\today
\end{flushleft}
\thispagestyle{empty}
\end{titlepage}}

\date{}
\firstpage{3118}{IC/95/34} {\huge \bf Flowing to four dimensions }
{E. Dudas$^{\,a,b,c}$ , C. Papineau$^{\,b,c}$ and V.A.~Rubakov$^{\,d}$}
{\small\sl $^a$   CERN Theory Division,
CH-1211 Geneva 23, Switzerland \\[-4mm]
\small\sl $^b$  CPhT $^\dagger$,
Ecole Polytechnique 91128 Palaiseau Cedex, France \\[-4mm]
\small\sl  $^c$ LPT $^\ddagger$,
B{\^a}t. 210, Univ. Paris-Sud, 91405 Orsay Cedex, France \\[-4mm]
\small\sl  $^d$ Institute for Nuclear Research of the Russian Academy of Sciences,\\[-4mm]
\small\sl 60th October Anniversary prospect 7a, Moscow 117312,
Russia} {We analyze the properties of a model with  four-dimensional
brane-localized Higgs type potential of a six dimensional scalar
field satisfying the Dirichlet boundary condition on the boundary of
a transverse two-dimensional compact space. The regularization of
the localized couplings generates  classical renormalization group
running. A tachyonic mass parameter grows in the infrared, in
analogy with the QCD gauge coupling in four dimensions. We find a
phase transition at a critical value of the bare mass parameter such
that the running mass parameter becomes large in the infrared
precisely at the compactification scale. Below the critical
coupling, the theory is in symmetric phase, whereas above it
spontaneous symmetry breaking occurs. Close to the phase transition
point there is a very light mode in the spectrum. The  massive
Kaluza-Klein spectrum at the critical coupling becomes independent
of the UV cutoff.} \setcounter{page}{0}

\section{Introduction}
 
Dynamical generation of small mass scales via logarithmic
renormalization group running and dimensional transmutation is a
common feature of many field theories. This phenomenon often occurs
in theories which, in the limit of vanishing couplings, possess
massless degrees of freedom. Once small couplings are turned on in
UV, a theory may become strongly coupled in IR, with the
corresponding IR scale serving as the mass scale of the low energy
effective theory. A notable example of such a case is of course QCD.

It is less common that massless or very light degrees of freedom
emerge upon turning on small couplings in a theory which originally
had heavy states only. 
Unless the existence of massless states is dictated by a symmetry
(e.g., Goldstone modes), this requires some sort of IR--UV mixing.
In this paper we present simple examples of 
this sort. In general terms, our models have two widely separated
high energy scales, the UV cutoff scale $\Lambda$ and the ``intermediate
scale'' $R^{-1}$ (the reason for the latter notation will become
clear shortly). Once the small coupling $\mu = \mu (\Lambda)$ is turned
on at the UV cutoff scale, it experiences renormalization group
running and increases towards low energies. There exists a (still
small) critical value $\mu(\Lambda) = \mu_c$ such that the running
coupling blows up precisely at the scale $R^{-1}$, i.e.,
$\mu(R^{-1}) = \infty$. At this point a massless mode appears in the
spectrum.

For $\mu $ slightly smaller than $\mu_c$, the light mode has
positive mass squared proportional to $(\mu_c^2 - \mu^2)$, whereas
at $\mu$ slightly above $\mu_c$ the low energy theory is in a
symmetry breaking phase. Thus, $\mu(\Lambda) = \mu_c$ is the point
of the second order phase transition.

These features are reminiscent of  phenomena appearing in string 
theory \cite{strong}, where
 strong coupling is associated with the presence of new light
degrees of freedom in the spectrum at particular points in the
moduli space. Our models are much simpler, and, in fact, the
features described above occur at the level of  classical field
theory. Specifically, we consider models which contain
six-dimensional scalar fields whose couplings are localized on a
four-dimensional brane. Their peculiar property, due to the
codimension-two nature of the couplings, is that the would-be mass
parameter $\mu^2$ is actually dimensionless and experiences a
classical (from the field theory viewpoint) running\footnote{The
classical running was recently discussed in connection with neutrino
masses in Ref.~\cite{dgv}. The phenomenon is well-known in two-dimensional
quantum mechanics, see e.g. Ref.~\cite{qm}.}, coming  from the regularization 
of UV
logarithmic divergences~\cite{gw,moz} 
(see also  Ref.~\cite{march}).
This regularization has a natural interpretation as
 finite transverse size
$\epsilon=\Lambda^{-1}$ of the brane. Remarkably enough, the
coupling becomes stronger in the infrared provided that the coupling
is negative or, equivalently, the 4d localized mass is tachyonic. We
consider the models on a disk in transverse dimensions, with the
Dirichlet boundary condition. The radius $R$ of the disk serves as
the intermediate scale. As described above, we find that the
critical value of the bare coupling, $\mu(\Lambda) = \mu_c$ (such
that $\mu(R^{-1}) = \infty$)
 is the point of
 the second order phase transition, at which there is a massless mode.

From the viewpoint of phenomenology with extra dimensions, our models generate
effective four-dimensional theories with small mass scales out of six-dimensional
theories whose energy scales (both the UV scale $\Lambda \equiv \epsilon^{-1}$
and intermediate scale $R^{-1}$)
may be very high --- say, of the order of the Planck mass. This separation of scales
is somewhat similar to what happens in warped models~\cite{rs,lykken},
but, unlike in the
latter, the Kaluza--Klein excitations in our models have very large masses.
In our models as they stand, the (bare) couplings are free parameters, so the choice
$\mu \approx \mu_c$ (and hence the theory 
near the phase transition point) 
requires
fine
tuning\footnote{Somewhat analogous fine tuning in theories with codimension two has been
discussed in Ref.~\cite{libanov}.}. It would be interesting to see if there exists a
dynamical mechanism driving the theory to the critical coupling (more generally, driving
the parameters to the point where $\mu (R^{-1}) = \infty$).

In Section 2 we present 
 the simplest model realizing this mechanism.
Section 3 contains a detailed six dimensional description of the
spectrum and interactions near the critical coupling. In Section 4
we show that the light mode present near the critical coupling has
standard four-dimensional mass and interactions at low energies. In
Section 5 we discuss another property of the model, namely, that the
massive Kaluza-Klein spectrum becomes independent of the UV cutoff
precisely at the critical coupling. Section 6 presents a
generalization of the simple model discussed previously, by adding a
brane scalar which mixes with the bulk scalar. We show that in this
case a new critical coupling arises, above which the scalar
potential appears destabilized. Section 7 shows  that the new
critical coupling, determined 
 by the renormalization group arguments in
Section 6, can again be understood in terms of a phase transition
(in a regularized brane setup and with appropriate boundary
conditions). We conclude with brief remarks and prospects for
further studies.
\section{The model}
The theory under consideration is a scalar theory 
in 6d space with two flat compact codimensions, with
no scalar potential in the bulk and Higgs-type scalar potential on a
brane situated at the origin of the compact space, described by the action\footnote{We  use the
convention $(+,-,-,-,-,-)$ for the metric.}
\bea
&& S \ = \ \int d^4 x d^2 y \ \biggl[ {1 \over 2} (\partial_M \phi)^2 \
- \  V_\delta (\phi) \biggr] \ , \nonumber \\
&& V_\delta (\phi) \ = \ \left(-\frac{\mu^2}{2}\phi^2 + \frac{\lambda}{4}
\phi^4 \right) \cdot \delta^2 ({ y})  \ . \label{1}
\eea
We resolve the singularity at ${ y}=0$ by introducing a disk $r < \epsilon$ supporting the potential,
\bea
&& V(\phi) \ = \ \frac{1}{\pi \epsilon^2}\left(-\frac{\mu^2}{2}\phi^2 +
\frac{\lambda}{4} \phi^4 \right)  \quad {\rm for} \ 0 < r < \epsilon \ ,  \label{potential} \\
&& V(\phi) \ = \ 0  \quad \quad \quad \quad \quad \quad \quad \quad \quad \quad
{\rm for} \ \epsilon  < r < R \ .  \nonumber
\eea
The parameter $\mu^2$ is dimensionless and may be considered a coupling constant.
We will assume $  \mu^2 \ll 1$.
Goldberger and Wise \cite{gw} argued that the coupling runs with
the energy scale $Q$, i.e.,
\be
\mu^2 (Q) = \frac{\mu^2}{1 + \frac{\mu^2}{2\pi}
\ln \frac{Q}{\Lambda}} \quad ,
\label{running}
\ee
where $\Lambda$ is the UV cutoff and
$ \mu^2 \equiv \mu^2(\Lambda)$
is the coupling constant entering the potential (\ref{potential});
we identify $   \Lambda = \epsilon^{-1}$.
Equation~(\ref{running}) implies that the coupling $\mu^2(Q)$ grows in the infrared.

We consider a theory on a disk of radius $R$, which is assumed to be small
(but $R \gg \epsilon$). Let us impose the Dirichlet boundary condition
\[
   \phi(r=R) \ = \ 0 \ .
\]
If the coupling vanishes, or, more generally, if
the running coupling at the compactification scale is small,
$\mu^2(R^{-1}) \ll 1$, the theory is fully 6-dimensional: there are no
zero or light modes, whereas the Kaluza--Klein states have masses
of order $R^{-1}$.

The question is what happens when $\mu^2 (R^{-1})$ hits the infrared
pole. In other words, what does this theory describe when the bare coupling $\mu^2$ is equal or
close to its critical value
\be
    \mu_c^2 \ = \ \frac{2\pi}{\ln \frac{R}{\epsilon}} \ .
\label{muc}
\ee
This is a question of the classical field theory, as the running
(\ref{running}) occurs at the classical field theory level.
Our claim is that very close to the critical coupling there
is a light mode, whose mass is proportional to
$ \mu_c^2 - \mu^2$ and vanishes just at the critical coupling.
At $\mu^2 > \mu_c^2$ this mode is tachyonic, and the field develops  vacuum
expectation value. Furthermore, the low energy
theory may be described fully in four-dimensional terms, and it is
a four-dimensional scalar theory with (without) the Higgs mechanism for
$\mu^2 > \mu_c^2$ ($\mu^2 < \mu_c^2$).
  This result can be understood qualitatively by noticing that the Dirichlet boundary condition
forces all modes to acquire a mass, in the spirit of the
Scherk--Schwarz mechanism \cite{ss}. Then from a naive 4d viewpoint,
the mass of a would-be zero mode has both the contribution due to
the boundary condition, and an additional contribution, of opposite
sign, coming from the localized tachyonic term, 
 \be m^{2}_{(0),{\rm
naive}} \ = \ {z_0^2 \over  R^2} \ - \ {\mu^2 \over \pi R^2} \ ,
\label{m1} \ee where $z_0$ is the first zero of the Bessel function
$J_0$ and the factor $1 / (\pi R^2)$ in the second term comes from
the KK expansion of the zero mode. Equation (\ref{m1}) indeed
predicts phase transition, but for $\mu_c^2 = (\pi z_0^2)  $. The
correct value, eq.~(\ref{muc}), however, contains an inverse
logarithmic factor coming from the running and the correct
expression for the mass is actually slightly more involved than the
simple guess (\ref{m1}). As follows from (\ref{muc}) and explained
in more detail in the following section, $\mu_c^2$ can (and
actually, for perturbative treatment, has to) be small, which
requires that $\ln ( R / \epsilon )$ is large. Thus, it is
legitimate to make use of the leading-log approximation and this is
what we are going to do. Incidentally, in the theory on a disk, the
formula for the running (\ref{running}) is also valid in the
leading-log approximation only.

\section{ Six-dimensional description of the phase transition}
\label{sec:2}

\subsection{At the critical coupling}
\label{sub:crit}

Let us see that at the critical coupling, there exists a massless
mode about the background $\phi =0$. In this background, the
massless mode obeys the free massless
equation away from the brane
\[
   \Delta^{(2)} \phi \ = \ 0 \; , \;\;\; r > \epsilon
\]
and a free equation with negative mass squared inside the resolved
brane,
\[
 \Delta^{(2)} \phi \ + \ \frac{\mu^2}{\pi \epsilon^2} \phi=
0 \; , \;\;\; r < \epsilon \ .
\]
The massless mode is in the s-wave, so
the solution inside the resolved brane is
\be
 \phi (r) \ = \ f_0 \ \left(
1  - \frac{\mu^2}{4\pi \epsilon^2}\cdot r^2 + \dots
\right) \; , \;\;\; \quad  r <\epsilon \ ,
\label{inside}
\ee
where $f_0$ is a small amplitude and dots stand for terms of higher order
in $\mu r$.
Note that the second term here is suppressed with respect to the
first term at all $r <\epsilon$, since $\mu^2$ is small.
Higher order terms are even more suppressed.

The solution outside the brane is
\be
  \phi(r) \ = \ a \ln \frac{R}{r} \; , \;\;\; r> \epsilon \ ,
\label{outside}
\ee
where we used the Dirichlet boundary condition at $r=R$.
Matching conditions for $\phi$ and $d \phi / dr $ at $r=\epsilon$
give
\bea
   a \ln \frac{R}{\epsilon} &=& f_0 \ ,
\nonumber \\
   a &=& \frac{\mu^2}{2\pi} f_0 \ . \nonumber
\eea
These are consistent right at the critical coupling, $\mu^2 = \mu_c^2$.
We conclude that
at this value of the coupling, there indeed exists a zero mass state.
\subsection{The mass of the light state near the critical coupling}
\label{sec:mass}

To  calculate the mass of the light state at $\mu^2$ close to (and slightly below)
$\mu_c^2$, still in the background $\phi=0$, one has to solve the
following equation
away from the brane
\be
   \Delta^{(2)} \phi \ + \ p^2 \phi \ = \ 0 \; , \;\;\; r > \epsilon \ ,
\label{massive}
\ee
where $p^2$ is the 4d mass. Inside the brane, the term with 4-momentum
in the field equation is tiny, so the solution still has the form
(\ref{inside}). For $p^2 \ll R^{-2}$, the s-wave solution to
eq.~(\ref{massive}) is
\be
\phi (r) \ = \ a \ \left[ \ln \frac{R}{r}
- \frac{p^2 r^2}{4} \ln  \frac{R}{r} + \frac{p^2}{4}
(R^2 - r^2) \right] \ .
\label{16}
\ee
The terms suppressed by higher powers of $p^2R^2$ and/or
$p^2r^2$ are irrelevant.

Now the matching conditions for $\phi$ and $d \phi / dr $ at
$r=\epsilon$ are
\bea
   f_0 &=& a \ \left(\ln \frac{R}{\epsilon} + \frac{p^2 R^2}{4}
\right) \ , \nonumber \\
    \frac{\mu^2}{2\pi} f_0 &=& a \ .
\eea
This gives for the four-dimensional mass squared
\be
 p^2 \equiv m_{(4)}^2 \ = \ \frac{4}{R^2} \left(
\frac{2\pi}{\mu^2} - \ln \frac{R}{\epsilon} \right) \ = \ {8 \pi
\over R^2} \ {1 \over \mu^2 (R^{-1})} \ , \label{eq19} \ee where
$\mu^2 (R^{-1})$ is the running coupling evaluated at the
compactification scale $R^{-1}$. Equivalently, we can write \be
m_{(4)}^2 = \frac{8 \pi}{R^2} \left( \frac{1}{\mu^2} -
\frac{1}{\mu^2_c} \right) \ = \   \frac{8 \pi}{R^2} \ \frac{\mu_c^2
- \mu^2}{\mu_c^4} \label{mass} \ . \ee (hereafter we consider the
linear order in ($\mu_c^2 - \mu^2$)). Notice that  the expression
(\ref{eq19}) for the   4d mass contains only the renormalized
coupling  $\mu^2 (R^{-1})$, and not the bare coupling  $\mu^2$
itself.  At the critical coupling, the light mode 
possesses  
scale invariance properties\footnote{For example, by using the quadratic part
of the effective action and the solution of the field equations 
(\ref{inside})~--~(\ref{outside}) , it can be shown that precisely
at the critical coupling $\int_0^R dr \ r \ T_M^M (r) = 0 $, where  
$T_{MN} = \partial_M \phi  \partial_N \phi - \eta_{MN} [ {1 \over 2}
 (\partial \phi)^2 -V (\phi) ]$ is the energy momentum tensor. Classically,
the quadratic part of the action is naively scale invariant for any
coupling $\mu$. Due to the regularization of the delta function, this
scale invariance is really valid only at the critical coupling, in
analogy with quantum field theories at the fixed points of the beta 
functions and anomalous dimensions.}.
\subsection{The Higgs expectation value}

At $\mu^2 > \mu_c^2$
the mass (\ref{mass}) becomes tachyonic, and the scalar field
develops a  vacuum expectation value. To obtain it, we have to solve the field
equation inside the brane,
\be
  \Delta^{(2)} \phi - \frac{\partial V}{\partial \phi}=
0 \; , \;\;\; r < \epsilon
\label{eqins}
\ee
and match the solution to the outside profile given by eq.~(\ref{outside}).
The field $\phi$ changes inside the brane only slightly,
so in eq.~(\ref{eqins}) one can use a constant value of
$V^\prime \equiv \frac{\partial V}{\partial \phi}$,
evaluated at $\phi(r=0)$.
Proceeding as in subsection 3.1,
one obtains the
inside solution
\be
  \phi \ = \ \phi_0 + \frac{1}{4} V^{\prime}(\phi_0)\cdot r^2 \ .
\label{in}
\ee
The second term here is indeed small compared to the first one
for the same reason as in subsection 3.1.

Matching this solution to (\ref{outside}), one
finds the equation\footnote{The latter equation
is in fact valid for arbitrary (but still small)
$\mu^2$ exceeding $\mu_c^2$,
and not only near the critical point. Thus, the result
(\ref{phi0}) is also valid for any $\mu^2$ above $\mu^2_c$.}
 for $\phi_0$,
\[
\phi_0 \ = \ - \frac{\epsilon^2}{2}  V^\prime (\phi_0)
\cdot \ln \frac{R}{\epsilon} \ .
\]
This gives
\[
1 \ = \ \frac{1}{2\pi} \ln \frac{R}{\epsilon} \cdot
(\mu^2 - \lambda \phi_0^2) \ .
\]
Recalling the definition of $\mu_c^2$ one gets finally
\be
  \phi_0^2 \ = \ \frac{\mu^2 - \mu_c^2}{\lambda} \ .
\label{phi0}
\ee
Note that the solution outside the brane is simply
\be
\phi_c (r) \ = \ \phi_0 \cdot
  \frac{\ln \frac{R}{r}}{\ln \frac{R}{\epsilon}}
\label{out2}
\ee
and that inside the brane the solution is approximately constant
and equal to $\phi_c = \phi_0$.

Let us calculate the mass of the light state (the ``Higgs boson'')
in the background
$\phi_c$. Writing $\phi (r,p) = \phi_c(r) + \xi(r,p)$, one has
in the inner region,
neglecting the term with four-momentum and recalling that
$\phi_c$ equals to $\phi_0$ up to small corrections,
\[
\Delta^{(2)} \xi - V^{\prime \prime}(\phi_0) \xi =0
\; , \;\;\; r<\epsilon \ .
\]
In the outer region, $\xi$ obeys  free
scalar equation, so the solution for $p^2 \ll R^{-2}$
still has the form (\ref{16}). Proceeding as in
subsection 3.2,
one obtains for the Higgs mass
\[
p^2 \equiv m_{\xi}^2
\ = \ \frac{8 \pi}{R^2} \ \frac{\mu_c^2 + \pi \epsilon^2 V^{\prime \prime}
(\phi_0) }{\mu_c^4} \ .
\]
With $\phi_0$ given by eq.~(\ref{phi0}) 
one obtains finally
\be
m_{\xi}^2
\ = \ \frac{16 \pi}{R^2}\cdot \frac{\mu^2 - \mu_c^2}{\mu_c^4} \ .
\label{mH}
\ee
This expression is valid for small $(\mu^2 - \mu_c^2)$.

Let us now extend the model by considering the scalar field in
some representation of a global symmetry group $G$. Then the vev
$\phi_c (r)$  breaks this symmetry, and 4d Goldstone bosons
 appear. The Goldstone
excitations
have the form
$
   \phi (x^\mu, r) = \mbox{exp}(i\pi^a (x) T^a) \phi_c(r)
$
where $T^a$ are broken generators of $G$ and $\pi^a(x^\mu)$ are massless
4d Goldstone fields. The low energy
effective 4d action for the Goldstone fields takes the form of
the standard sigma-model action,
\[
S_{eff} \ = \ \sigma^2 \int~d^4x~\frac{1}{4} \left[
(\partial_\mu \pi^a)^2 +  \pi \cdot \partial_\mu \pi
\cdot \partial^\mu \pi + \dots \right] \ ,
\]
where the effective 4d Higgs vev is
\[
 \sigma^2 \ = \ \int~d^2y~\phi_c^2 \ .
\]
The contribution to this integral mainly comes from the outside region $ \epsilon < r < R$.
By using eq.~(\ref{out2}), one obtains
\bea
\sigma^2 &=& 2\pi \int_0^R ~rdr  \left(\phi_0 \cdot
  \frac{\ln \frac{R}{r}}{\ln \frac{R}{\epsilon}}\right)^2
\nonumber \\
&=&
\frac{\pi}{2} \ \frac{\phi_0^2 R^2}{\ln^2 (\frac{R}{\epsilon})} \quad = \quad
\frac{R^2}{8\pi \lambda} \cdot \mu_c^4 \ (\mu^2-\mu_c^2) \ .
\label{sigma2}
\eea
If there are gauge fields in the bulk (with Neumann boundary condition
for $A_\mu$ at $r=R$), then $\phi$ plays the role of the Higgs field.
The  light modes of
the gauge fields are constant in extra dimensions, so that
they obtain the masses $M_V^2 = g_{(4)}^2 \sigma^2$, where
$ g_{(4)}^2$ is the 4d gauge coupling and the effective 4d Higgs vev
is given precisely by eq.~(\ref{sigma2}). Notice that $M_V^2/m_{\xi}^2 >> 1$ if $\lambda \sim 1 / \Lambda^4$, whereas
the more interesting result $M_V^2/m_{\xi}^2 \leq 1$ is valid if $\lambda \geq R^4 \mu_c^8$.
\section{ Four-dimensional effective theory}
Let us see how the results of section~3 compare to the standard 4d
effective low energy theory. It is clear from
eqs.~(\ref{outside}), (\ref{16}), (\ref{out2})
 that up to small corrections, the interesting
field configuration away from the brane is
\be
    \phi(x^\mu, r) \ = \ \sigma (x^\mu) \cdot \zeta (r) \ ,
\label{eff1}
\ee
where
\[
  \zeta(r) \ = \ \sqrt{\frac{2}{\pi R^2}} \ \ln \frac{R}{r} \ .
\]
Resolving the brane is equivalent to defining
\bea
  \zeta (0) \ = \ \zeta (\epsilon) &=&  \sqrt{\frac{2}{\pi R^2}}  \ \ln
\frac{R}{\epsilon}
\nonumber \\
&=& \frac{2 \sqrt{2\pi}}{\mu_c^2 R} \ .
\label{eff2}
\eea
We are interested in the effective theory of the 4d field $\sigma (x^\mu)$.
The wave function $\zeta$ is normalized to unity,
\[
  \int~d^2y~ \zeta^2 \ = \ 1 \
\]
and therefore $\sigma (x)$ has a canonical kinetic term.
Its effective 4d potential is obtained by plugging (\ref{eff1})
and (\ref{eff2}) into the potential (\ref{1}) and the transverse
kinetic energy term. One finds
\[
  V_{eff} (\sigma) \ = \ \frac{m_{(4)}^2}{2} \sigma^2
+ \frac{\lambda_{(4)}}{4} \sigma^4 \ ,
\]
where
\[
\lambda_{(4)} = \lambda [\zeta(0)]^4
\ = \  \frac{64 \pi^2}{\mu_c^8} \ \frac{\lambda}{R^4}
\]
and
\[
m_{(4)}^2 \ = \ - \mu^2 [\zeta(0)]^2
+ 2\pi \int_{\epsilon}^R ~rdr~ (\zeta^\prime)^2 \ .
\]
The value of the latter is precisely the same as in (\ref{mass}),
which establishes the correspondence between the 6d and the 4d approaches
in the unbroken phase. For $m_{(4)}^2 <0$, the 4d expressions for
the vev and the Higgs mass are $
  \sigma^2  =  - \frac{m_{(4)}^2}{\lambda_{(4)}}
$
and
$
 m_\xi^2 =  - 2 m_{(4)}^2
$.
These coincide with (\ref{sigma2}) and (\ref{mH}), respectively,
so that the correspondence exists in the broken phase as well.

\section{Massive spectrum near the critical coupling}

Let us now work out the massive spectrum at and slightly below
the critical coupling.
In this case the background is $\phi = 0$ and by defining
the wave functions
$
\phi (r ,\theta) \ = \ e^{i l \theta} \ \chi_l (r) \ , \label{m01}
$
 we get the Schr\"odinger equations
\bea
&&  \chi^{''}_l + {1 \over r} \ \chi'_l +
(p^2 + {\mu^2 \over \pi \epsilon^2}
- {l^2 \over r^2}) \ \chi_l  \ = \ 0 \quad ,\quad {\rm for}  \ r < \epsilon
\ , \nonumber \\
&&  \chi^{''}_l + {1 \over r} \ \chi'_l + (p^2 - {l^2 \over r^2} ) \ \chi_l
\ = \ 0 \quad ,\quad \quad \quad \quad {\rm for}  \ r > \epsilon
\ . \label{m2}
\eea
The solutions for the positive mass squared (positive $p^2$) wave functions,
satisfying the Dirichlet boundary
condition at $r = R$ are
\bea
&& \chi_l (r) \ = \ A \ J_l \biggl(\sqrt{p^2 + {\mu^2 \over \pi \epsilon^2}} r
\biggr)
\quad , \quad \quad \quad \quad \quad {\rm for} \ r < \epsilon  \ , \nonumber \\
&& \chi_l (r) \ = \ B \ \biggl[ J_l (p r) - {J_l (p R) \over N_l (pR)}
 N_l (p r)  \biggr] \quad , \quad {\rm for} \ r > \epsilon  \ ,
\ \label{m3} \eea where $J_l$ and $N_l$ are the Bessel functions of
the first and second kind. The matching condition of the logarithmic
derivative of the wave function, which defines the massive spectrum,
is
\[
\sqrt {p^2 + {\mu^2 \over \pi \epsilon^2}} \ \ { J'_l \biggl(
\sqrt {p^2 \epsilon^2 + {\mu^2 \over \pi }}  \biggr) \over
 J_l \biggl( \sqrt {p^2 \epsilon^2 + {\mu^2 \over \pi }}  \biggr)}
\ = \
p \ \frac{ J'_l ( p \epsilon ) -  {J_l ( p R ) \over N_l ( p R )}
 N'_l ( p \epsilon ) }{
 J_l ( p \epsilon) -  {J_l ( p R ) \over N_l ( p R )}
 N_l ( p \epsilon )} \ . \label{m4}
\]
In the following we concentrate on the s-wave solutions $l=0$ near the
critical coupling $\mu_c$. Under the physically sensible condition
$p \epsilon << 1$, for  
perturbative coupling at the cutoff
$\mu^2 << 1$ and by using the expansion of the Bessel functions for small
argument,
\[
J_0 (z ) \simeq 1 - {z^2 \over 4} \quad, \quad N_0 (z) \simeq {2 \over \pi}
\ \ln {z \over 2} \ , \label{m5}
\]
 we 
find at the critical coupling $\mu = \mu_c$ the eigenvalue equation
determining the massive spectrum
\be
{2 \over \pi} \ {J_0 (p R) \over N_0 (p R)} \ \ln{p R \over 2} \ =
\ 1 \ , \label{m6}
\ee
up to negligibly small terms of order $p^2 \epsilon ^2$. Interestingly enough,
precisely at the critical coupling, the UV cutoff dependence of the KK
masses cancels  out from eq.~(\ref{m6}). If we redo the same
analysis slightly below the critical coupling,
 we find
\be
{2 \over \pi} \ {J_0 (p R) \over N_0 (p R)} \ \biggl( \ln{p R
\over 2} + \frac{\mu_c^2 - \mu^2}{\mu_c^2}   \ln{R \over \epsilon} \biggr) \ = \ 1 \ ,
\label{m6a}
\ee
and the KK masses start to depend on the cutoff.

In analogy to the lightest mass, the KK masses (\ref{m6a}) 
can be expressed
as functions of the running coupling $\mu^2 (p)$ evaluated at the pole.
Indeed, eq.~(\ref{m6a}) can be written as
\be {J_0 (pR) \over N_0 (pR)} \ = \ {1 \over 4} \ \mu^2 (p) \ , \ee
where the running coupling is given by 
eq.~(\ref{running}). 

We note in passing that keeping subleading terms in the
expansion of the Bessel
functions,
one obtains, near the critical coupling,
the following equation instead of eq.~(\ref{m6a}),
 \[
 {2 \over \pi} \ {J_0 (p R) \over N_0 (p R)} \ \biggl( \ln{p R \over 2} +\frac{3}{16} +
\frac{2\pi (\mu_c^2 - \mu^2) }{\mu_c^4} \biggr) \ = \ 1 \ .
\]
  We then define the corrected value for the critical coupling
$\hat{\mu}_c$ as
\[ \hat{\mu}_c^2 = \mu_c^2 - \frac{3}{32\pi}
\mu_c^4 \ ,
\]
 where $\mu_c$ by definition is given by eq.~(\ref{muc}).
Clearly, the second term here is subleading; it is suppressed, as
compared to the first term, by $[\ln(R/\epsilon)]^{-1}$. After this
qualification, eq.~(\ref{m6a}) remains valid, with $\hat{\mu}_c$
substituted for $\mu_c$.
Formulae
in previous sections remain valid too, again with $\hat{\mu}_c$ used
instead of $\mu_c$ everywhere. In the leading-log approximation one
neglects the difference between $\hat{\mu}_c$ and $\mu_c$.
Thus, consistent leading-log approximation indeed
corresponds to neglecting
higher order terms in the expansion of $J_0$ and its
derivatives.

It is likely that other physical quantities are also
cutoff-independent at the critical coupling. For example, by using
the results of Goldberger and Wise~\cite{gw}, we  find the scalar
propagator at the critical point. It was shown in Ref.~\cite{gw}
that the (Euclidean) Dyson resummation of the scalar propagator, by
including the brane localized mass insertions, in a mixed
representation, 4d momentum space and 2d extra-dimensional
coordinate space, is (cf. Ref.~\cite{dgp})
\begin{eqnarray}
 G (p, y , y^\prime )
 & = &   D (p,y , y^\prime) + \mu^2 \ D
(p,y  , 0 )  D (p, 0 ,  y^\prime)
 \nonumber
\\
&&
\ \ +
\mu^4 \   D (p,y , 0) \ D (p, 0,0)   D
(p, 0 ,  y^\prime) +  \cdots \ \nonumber
\\
& = &  D (p,y ,  y^\prime)  + {\mu^2 \over 1 - \mu^2
D (p,0, 0)} \
 D (p, y ,  0)  D (p, 0 ,  y^\prime) \ , \
    \label{rd1}
\end{eqnarray}
where $D (p,y ,  y^\prime)$ is the free scalar six-dimensional
propagator.
For the scalar field with the
Dirichlet boundary condition at $r=R$, the propagator at the origin of the compact space $D
(p, 0, 0)$ can be approximately evaluated, for different
values of the four-dimensional momentum, as
\begin{eqnarray}
&& D (p, 0,  0)  \sim {1 \over 4 \pi} \ln {\Lambda^2 \over
p^2} \quad , \quad \quad \quad
\quad \quad \quad \quad \; {\rm for} \ \ p^2 >> 1/R^2 \ , 
\nonumber
\\
&& D (p, 0, 0)  \sim {1 \over 4 \pi} \ln (\Lambda^2 R^2)  + O (p^2 R^2) \quad , \quad {\rm for} \ \ p^2 << 1/R^2 \
. \label{rd04}
\end{eqnarray}
Note that for $p^2 \gg 1/R^2$ the propagator (\ref{rd1}) indeed
exhibits logarithmic scale-dependence typical of  renormalization
group running of the coupling $\mu^2$. It can then be easily shown
that at the critical coupling $\mu = \mu_c$, \bea && {\mu_c^2 \over
1 - \mu_c^2 D (p, 0 , 0)} \ = \ {2 \pi \over \ln (pR)} \ ,
\quad \;
\ {\rm for}
\ \ p^2 \ >> \ {1 \over R^2}  \ , \ {\rm and} \  \nonumber \\
&&  {\mu_c^2 \over 1 - \mu_c^2 D (p, 0 , 0)} \sim O \left({1 \over p^2 R^2 }\right) \   \ , \
{\rm for} \ \ p^2 \ << \ {1 \over R^2} \ .
\eea
These quantities are indeed independent of the UV cutoff. The existence of the
massless mode is again obvious: it is
seen as a pole in the propagator at $p^2=0$
for the critical coupling $\mu=\mu_c$.

\section{Mixing with  brane scalar: classical running and critical
couplings}

\subsection{Running couplings}

There are several reasons to try to generalize the previous 6d toy
model by including brane-localized fields. In particular, we would
like to better understand the possible role of brane fields in the
value of the critical coupling(s) and also the dynamics of the field
theory living on the brane close to the critical couplings. The
next-to-simplest example includes, in addition to the bulk scalar
$\phi$, a brane-localized scalar $H$ mixing with $\phi$. The
corresponding action is  
\bea
  S = \int d^4 x d^2 y && \left(
\frac{1}{2}(\partial_A \phi)^2 - \delta^2 (y) \left[ - \frac{\mu^2}{2} \phi^2
 + {\lambda \over 4}   \phi^4  \right. \right. \nonumber \\
&& \left. \left.- \frac{1}{2}(\partial_{\mu} H)^2 + {h \over 2} \phi H^2
+ {\lambda' \over 4}  H^4  \right] \ \right)  \ .
\label{mix1}
\eea
The brane scalar $H$ 
is massless in the vacuum $\langle \phi \rangle
=0$. Notice that $\mu^2$, $h$ and $\lambda'$ are dimensionless
couplings and can naturally mix, whereas $\lambda$ is dimensionfull
and does not mix. It is straightforward to work out the classical
(tree-level) field theory diagrams which contribute to the running
of the coupling constants. We find the RG equations \bea
&& Q {d \mu^2 \over d Q} = - {1 \over 2 \pi} \mu^4 \ , \nonumber \\
&& Q {d h \over d Q} = - {1 \over 2 \pi} h \mu^2 \ , \nonumber \\
&& Q {d \lambda' \over d Q} = + {1 \over 4 \pi} h^2 \ .
\nonumber
\eea
The integration of these RG equations gives
\bea
&&
\mu^2(Q) = \frac{ \mu^2}{1 +  {\mu^2 \over 2 \pi} \ln {Q \over \Lambda}}
\ , \nonumber \\
&& h (Q) = \frac{ h}{1 +  {\mu^2 \over 2 \pi} \ln {Q \over \Lambda}}
\ , \nonumber \\
&&\lambda' (Q) = \lambda' + {h^2 \over 4 \pi} \frac{ \ln
{Q \over \Lambda}} {1 +  {\mu^2 \over 2 \pi} \ln {Q \over \Lambda }}
\ ,
\label{mix3}
\eea
where $\mu \equiv \mu (\Lambda)$, etc.
The running couplings $\mu^2(Q)$ and $h(Q)$ become strong at the critical coupling
$\mu_c$ defined in the previous sections. Notice, however, that the
coupling $\lambda' (Q)$ becomes negative and appears to destabilize
the potential
for smaller values of the bare coupling $\mu$; actually, $\lambda'
\rightarrow - \infty$ at the critical coupling! As we will see in a
moment, of particular relevance for this problem is the combination
of couplings ${\bar \mu}^2 \equiv \mu^2 + h^2 / (2 \lambda')$, which
runs according to
\be
{1 \over [\mu^2 + h^2 / (2 \lambda')](Q)} = {1
\over \mu^2 + h^2 / (2 \lambda')} +  {1 \over 2 \pi} \ln
{Q \over \Lambda } \ .
\label{mix4}
\ee
Clearly we find here a {\it second critical coupling} defined by
\be
{1 \over 2 \pi} [ \mu^2 +
h^2 / (2 \lambda')]_c \ \ln {R \over \epsilon}
 = 1 \ .
\label{mix5}
\ee
Remarkably, it occurs at $\mu < \mu_c$ (we continue to use the notation
$\mu_c$ for the first critical coupling). At the new critical coupling,
$\mu^2 (R^{-1})$ and  $h (R^{-1})$ are finite, whereas
\[
\lambda'(R^{-1}) \ = \ 0 \ .
\label{mix6}
\]
The new critical coupling (\ref{mix5}) is therefore the one above
which the scalar potential appears destabilized.


 We now turn to  closer examination of the new critical
coupling, expectation values of the fields, mass spectrum
and the effective
theory associated to it.

\subsection{Simplified description near the new critical coupling}

A simple but heuristic description near the new critical coupling
is to regularize the delta function as above and write a 4d
kinetic term for $H$ inside the disk $r < \epsilon$. This would be
consistent if the field equations implied that $H$ is independent of
$r$ inside the resolved brane.
We will see in the next section that with regularized $\delta$-function,
this actually is not the case. 
We will turn to the accurate treatment
in the next section, and here we proceed with the somewhat
heuristic analysis.

In this
description, for $r < \epsilon $ the scalar potential is
\be
V(r) = {1 \over \pi \epsilon^2} \biggl(- \frac{\mu^2}{2} \phi^2  +
{h \over 2} \phi H^2 + {\lambda' \over 4}  H^4
+ {\lambda \over 4}  \phi^4  \biggr) \ ,
\label{s1}
\ee
whereas $V  = 0$ for $r > \epsilon \ $.  The classical field equations close
to the critical coupling, by keeping only the
relevant terms, are
\[
 \Delta^{(2)} \phi \ = \ 0 \;\;\; \ ,  \quad \quad \quad  
r > \epsilon \ ,
\]
and
\bea
 \Delta^{(2)} \phi \ + \ \frac{\mu^2}{\pi \epsilon^2} \phi \ - \ {h \over
2 \pi \epsilon^2} H^2 \ &=& \ 0 \; ,   \nonumber \\
  h \phi H \ + \ \lambda' H^3  \ &=& \ 0 \;\;\; , \  \quad \quad \quad  
r < \epsilon \ .
\label{s2}
\eea
By  combining the field equations, we find at $r<\epsilon$
\[
\pi \epsilon^2 \Delta^{(2)} \phi \ + \ \left(\mu^2 +
{h^2 \over 2 \lambda'}\right) \phi  \ = \ 0 \ . \label{s3}
\]
The analysis now is similar to that in Section 3, but the role of the coupling is
played by
the combination ${\bar \mu}^2 \equiv  \mu^2 + ({h^2 / 2 \lambda'})$.
This leads precisely to the condition (\ref{mix5}) for the second
critical coupling.
In complete analogy to the
analysis in Section 3, we find that there is a light state whose
mass close to the second critical coupling and $\bar{\mu} \leq \mu_c$  is
 \be
p^2 \ = \ M^2 \ = \ {8 \pi \over R^2} \ {\mu_c^2 - {\bar \mu}^2 \over
\mu_c^4} \ .
\label{s5}
 \ee
 The scalar expectation values for ${\bar \mu} \geq \mu_c
$ are given by
\be
\phi_0^2 \ = \ {{\bar \mu}^2 -  \mu_c^2 \over
\lambda} \quad , \quad H_0^2 = - \ {h \over \lambda'} \ \phi_0 \ .
\label{s6}
\ee
Hence, the second critical coupling is the point of the second order phase
transition, at which both $\phi$ and $H$ obtain expectation values.
\section{Resolving the delta-function in the model with  brane scalar}

\subsection{Regularizing the delta-function}

Let us now regularize the delta-function more carefully.
We include a gradient term for $H$ in the two
extra dimensions to account for the radial variation of $H$
inside the resolved brane. 
As before, the brane is
a disk of radius $\epsilon$. The action inside the brane ($r<\epsilon$) 
is 
\bea
S_{in} &=& \int~d^2 y~d^4 x \left[\left ( \frac{1}{2}
(\partial_A \phi)^2 + \frac{\mu^2}{2\pi \epsilon^2}  \phi^2  \right) \right.
\nonumber \\
&+& \left.  \frac{1}{\pi \epsilon^2} \left(
 \frac{1}{2} (\partial_\mu H)^2  - \frac{1}{2}
(\partial_a H)^2 + \frac{m^2}{2} H^2 -
\frac{1}{2} h \phi H^2  -
\frac{\hat{\lambda}^\prime}{4} H^4 \right) \right] \ .
\nonumber
\eea
We are going to discuss the critical couplings, so the self-coupling
of $\phi$ is unimportant (see below) and is neglected here. Since we
interpret $H$ as a brane field, its boundary condition at
$r=\epsilon$ is Dirichlet. In order that 
the massless excitation of $H$ above  
the background  $ H =0$ be possible, we 
introduced a mass term  $m^2 \sim \epsilon^{-2}$. 
This is part of our regularization
procedure.

Note the ``wrong'', tachyonic sign of $m^2$ and also of $\mu^2$. Note also that
we introduced the coupling $\hat{\lambda}^\prime$, which is not yet quite the
quartic
coupling constant in the delta-function limit.

Let us consider the linearized equation for $H$ in the background
$\phi=H=0$, 
\be
 \Delta^{(2)} H + m^2 H \equiv LH = - p^2 H \ . 
\label{linearH}
\ee
We choose $m^2$ in such a way that this equation has a zero mode $p^2=0$
with the Dirichlet boundary condition at $r=\epsilon$ (we will not need the explicit
expression for $m^2$ or the zero mode). Let
\[
        H^{(0)} = H^{(0)} (r/\epsilon)
\]
denote the zero mode, normalized in such a way  that
\be
2  \int^1_0~\rho
d\rho \ [H^{(0)}(\rho)]^2 \ = \ 1 \ ,
\label{79}
\ee
where $\rho=r/\epsilon$. The low
energy effective theory for the brane field $H$ is obtained by
writing
\[
   H(x,y) \ = \ H_B (x) \ H^{(0)} (y) \ .
\]
Naively at least, this results in a 
brane field $H_B (x^\mu)$ 
with canonically
normalized kinetic term, a $\delta$-function interaction of $H_B$
with $\phi$ with coupling constant $h$, and a quartic self-interaction
of $H_B$ with coupling constant
\be
  \lambda^\prime = 2  \hat{\lambda}^\prime \int_0^1~d\rho \rho \ [H^{(0)}(\rho)]^4  \ .
\label{v4*}
\ee
This coupling differs from $\hat{\lambda}^\prime$ by a numerical factor of
order 1.
\subsection{ Phase transition at the second critical coupling}

Let us now show that at the second critical coupling, the second order phase transition
indeed occurs, so that above this point the fields $\phi$ and $H$ develop
expectation values. A signal for the phase transition is the existence of an approximate
modulus near the origin in the field space, i.e., the direction along which
the curvature of the effective scalar potential vanishes. In other words, at the phase transition
point there exists a non-trivial 
 $x^\mu$-independent solution to the field equations with small
(but otherwise arbitrary) amplitude. To see that this is indeed the case precisely at the second
critical coupling, we neglect $\lambda \phi^4$ term in the action and write the
field equations inside the resolved brane for the fields $\phi$ and $H$ 
depending on $r$ only, 
%
\bea
   \Delta^{(2)} H + m^2 H -h \phi H
-\hat{\lambda}^\prime  H^3 &=& 0 \ , 
\label{cl1} \  
\\
    \pi \epsilon^2 \Delta^{(2)} \phi
 + \mu^2 \phi -\frac{h}{2}  H^2 &=& 0 \quad {\rm for} \ r < \epsilon \ .
\label{cl2}
\eea
Outside the brane, the field $\phi$ still obeys the 2d Laplace equation.

Let us begin with eq.~(\ref{cl1}). This equation has
the form
\be
 L \ H \ = \ J \ ,
\label{LH}
\ee
where
\[
  J \ = \ h  \phi H
\ + \ \hat{\lambda}^\prime  H^3
\]
and the operator $L$ is defined in (\ref{linearH}).
Since $L$ has a zero mode $H^{(0)} (y)$, eq.~(\ref{LH}) has a solution
(with the Dirichlet condition at $r=\epsilon$) if and only if $J$
is orthogonal to the zero mode,
\be
\int_0^1 ~\rho d\rho \ H^{(0)} \ J  \ = \ 0 \ .
\label{constraint}
\ee
If this property is satisfied, one can solve eq.~(\ref{LH})
perturbatively, considering $J$ as perturbation (given that
$\Delta^{(2)}$ and $m^2$ are of order $\epsilon^{-2}$
inside the resolved brane). 
To the zeroth order one has
\be
  H(y) \ = \ H_0 \ H^{(0)} (y) \ ,
\label{lo}
\ee
where $H_0$ is a constant.

Let us now turn to eq.~(\ref{cl2}). Recalling that the boundary condition
for $\phi$ at $r=\epsilon$ is
\be
    \frac{d\phi /d\rho}{\phi} (\rho=1) \ = \ - \frac{1}{\ln \frac{R}{\epsilon}}
\label{junction} \ ,
\ee
one observes that $\phi$ is ``large'' (inside
the brane) compared to $d\phi / d\rho$. Hence we write the solution
by considering the last two terms in (\ref{cl2}) 
as 
perturbations:
\[
\phi (\rho) = \phi_0 + \frac{h}{2 \pi}H_0^2
\int_0^\rho \frac{d\rho^{\prime}}{\rho^\prime} \int_0^{\rho^\prime}
\rho^{\prime \prime} d\rho^{\prime \prime} [H^{(0)}(\rho^{\prime
\prime})]^2 - \frac{\mu^2}{4 \pi}\rho^2 \phi_0 + \dots \ ,
\]
where the
leading order expression for $H(y)$, eq.~(\ref{lo}) has been used.
Since $h$ and $\mu$ are small, the second and third terms here are
indeed small compared to the first term on the right hand side.

In view of  eq.~(\ref{79}), 
the boundary condition (\ref{junction})
gives 
\be
  \frac{1}{2\pi}\left( \frac{h}{2} H_0^2 - \mu^2 \phi_0 \right) =
- \frac{1}{\ln \frac{R}{\epsilon}} \phi_0 \ . 
\label{v2*}
\ee
To find another relation between $\phi_0$ and $H_0$, we use
eq.~(\ref{constraint}). We insert the leading order expressions for
$\phi$ (that is $\phi = \phi_0$) and $H$ (that is $H \ = \ H_0 H^{(0)}(y)$) and
obtain
\be
h \phi_0 H_0 + \lambda^\prime H_0^3 =0 \ ,
\label{v2**}
\ee
where $\lambda^\prime$ is given by eq.~(\ref{v4*}).
Equations (\ref{v2*}) and (\ref{v2**}) are consistent precisely at the second
critical coupling defined by eq.~(\ref{mix5}). Thus, we find that
at the second critical coupling,
there 
 exists an approximate modulus near the origin of the field space.
The second critical coupling is indeed the 
 point of the second order
phase transition.

The rest of the analysis parallels that of Section 3. With $\lambda \phi^4$ term
included, the expectation values of the fields just above the critical coupling
are indeed given by eq.~(\ref{s6}), while the new light state has mass (\ref{s5}).
We conclude that our more accurate analysis confirms the results of the
heuristic treatment of subsection~6.2.

\section{Conclusions}

The toy models we analyzed in this paper possess 
 fairly rich
physics: running couplings, phase transitions, spontaneous symmetry
breaking and infrared strong dynamics which all occur at the level
of classical field theory.
In addition of being an interesting laboratory for studying difficult physics issues such
as the ones just mentioned,
our models could be of some relevance for phenomenological string or extra dimensional models with
small radii and large (close to the Planck mass) fundamental mass scale. Indeed, if for some
(eventually dynamical) reason 
 the microscopic parameters are very close to the point (surface)
defining the critical coupling(s), very light modes and standard four dimensional physics are generated
out of a higher dimensional theory with  large scales only. 
This mechanism could be of  
relevance
for the problem of electroweak symmetry breaking and mass generation or/and for
explaining the smallness of supersymmetry breaking in appropriate supersymmetric extensions.
Of course, an important step to make before addressing these phenomenological issues is the
inclusion of other fields like chiral 4d fermions and gravitational interaction.

\vskip 16pt
\begin{flushleft}
{\large \bf Acknowledgments}
\end{flushleft}

\noindent It is a pleasure to thank T.~Gherghetta, D.~Gorbunov, P.~Hosteins,
D.~Levkov, M.~Libanov, J.~Mourad, S.~Sibiryakov,
A. Vainshtein and S.K.~Vempati for useful discussions.
This work was supported in part by the CNRS PICS no. 2530 and 3059, INTAS grant
03-51-6346, the RTN grants MRTN-CT-2004-503369, MRTN-CT-2004-005104,
a European Union Excellence Grant, MEXT-CT-2003-509661 and RFBR grant 05-02-17363a.


 \end{document}


\bibitem{extra}
 V.~A.~Rubakov and M.~E.~Shaposhnikov,
  Phys.\ Lett.\ B {\bf 125} (1983) 139
and
  Phys.\ Lett.\ B {\bf 125} (1983) 136;
 K.~Akama,
  Lect.\ Notes Phys.\  {\bf 176} (1982) 267
  [arXiv:hep-th/0001113];
M.~Visser,
  Phys.\ Lett.\ B {\bf 159} (1985) 22
  [arXiv:hep-th/9910093];
 I.~Antoniadis,
  Phys.\ Lett.\ B {\bf 246} (1990) 377.

\bibitem{add}
N.~Arkani-Hamed, S.~Dimopoulos and G.~R.~Dvali,
  Phys.\ Lett.\ B {\bf 429} (1998) 263
  [arXiv:hep-ph/9803315] and
  Phys.\ Rev.\ D {\bf 59} (1999) 086004
  [arXiv:hep-ph/9807344];
 I.~Antoniadis, N.~Arkani-Hamed, S.~Dimopoulos and G.~R.~Dvali,
  Phys.\ Lett.\ B {\bf 436} (1998) 257
  [arXiv:hep-ph/9804398].

\bibitem{ddg}
K.~R.~Dienes, E.~Dudas and T.~Gherghetta,
  Phys.\ Lett.\ B {\bf 436} (1998) 55
  [arXiv:hep-ph/9803466] and
  Nucl.\ Phys.\ B {\bf 537} (1999) 47
  [arXiv:hep-ph/9806292].

\bibitem{march}
  N.~Arkani-Hamed, S.~Dimopoulos and J.~March-Russell,
  arXiv:hep-th/9908146.

\bibitem{Randall:2001gc}
  L.~Randall and M.~D.~Schwartz,
  Phys.\ Rev.\ Lett.\  {\bf 88} (2002) 081801
  [arXiv:hep-th/0108115].

\bibitem{Agashe:2002pr}
  K.~Agashe, A.~Delgado and R.~Sundrum,
  Annals Phys.\  {\bf 304} (2003) 145
  [arXiv:hep-ph/0212028].

\bibitem{rs}
L.~Randall and R.~Sundrum,
  Phys.\ Rev.\ Lett.\  {\bf 83} (1999) 3370
  [arXiv:hep-ph/9905221] and
  Phys.\ Rev.\ Lett.\  {\bf 83} (1999) 4690
  [arXiv:hep-th/9906064].

\bibitem{ggh}
  H.~Georgi, A.~K.~Grant and G.~Hailu,
  Phys.\ Lett.\ B {\bf 506} (2001) 207
  [arXiv:hep-ph/0012379].

\bibitem{gw}
  W.~D.~Goldberger and M.~B.~Wise,
  Phys.\ Rev.\ D {\bf 65} (2002) 025011
  [arXiv:hep-th/0104170].

\bibitem{dgv}
E.~Dudas, C.~Grojean and S.~K.~Vempati,
  arXiv:hep-ph/0511001.

\bibitem{ab}
C.~P.~Bachas,
  JHEP {\bf 9811} (1998) 023
  [arXiv:hep-ph/9807415];
  I.~Antoniadis and C.~Bachas,
  Phys.\ Lett.\ B {\bf 450} (1999) 83
  [arXiv:hep-th/9812093];
 I.~Antoniadis, C.~Bachas and E.~Dudas,
  Nucl.\ Phys.\ B {\bf 560} (1999) 93
  [arXiv:hep-th/9906039];
N.~Arkani-Hamed, S.~Dimopoulos and J.~March-Russell,
  arXiv:hep-th/9908146.

\bibitem{ss}
 J.~Scherk and J.~H.~Schwarz,
  Nucl.\ Phys.\ B {\bf 153} (1979) 61;
E.~Cremmer, J.~Scherk and J.~H.~Schwarz,
  Phys.\ Lett.\ B {\bf 84} (1979) 83.

\bibitem{strong}
  A.~Sagnotti,
  Phys.\ Lett.\ B {\bf 294} (1992) 196
  [arXiv:hep-th/9210127];
N.~Seiberg and E.~Witten,
  Nucl.\ Phys.\ B {\bf 471} (1996) 121
  [arXiv:hep-th/9603003].

\bibitem{six}
 S.~M.~Carroll and M.~M.~Guica,
  arXiv:hep-th/0302067;
I.~Navarro,
  JCAP {\bf 0309} (2003) 004
  [arXiv:hep-th/0302129];
Y.~Aghababaie, C.~P.~Burgess, S.~L.~Parameswaran and F.~Quevedo,
  Nucl.\ Phys.\ B {\bf 680} (2004) 389
  [arXiv:hep-th/0304256];
 S.~Randjbar-Daemi and V.~A.~Rubakov,
  JHEP {\bf 0410} (2004) 054
  [arXiv:hep-th/0407176].

\end{thebibliography}

 \end{document}